\documentclass[%
 aip,
 amsmath,amssymb,
 reprint,%
]{revtex4-1}

\usepackage{graphicx}
\usepackage{dcolumn}
\usepackage{bm}

\usepackage[utf8]{inputenc}
\usepackage[T1]{fontenc}
\usepackage{mathptmx}
\usepackage{etoolbox}
\usepackage{hyperref}

\usepackage{xcolor}

\makeatletter
\def\@email#1#2{%
 \endgroup
 \patchcmd{\titleblock@produce}
  {\frontmatter@RRAPformat}
  {\frontmatter@RRAPformat{\produce@RRAP{*#1\href{mailto:#2}{#2}}}\frontmatter@RRAPformat}
  {}{}
}%
\makeatother
\begin{document}

\preprint{AIP/123-QED}

\title[Probing Non-Equilibrium Baths]{
Probing Non-equilibrium baths:
Frequency-Resolved Thermometry and Quantum Heat Current Turnover
}
\author{Akhil Bhartiya}
\author{Tobias Kramer}%
\affiliation{ 
Institute for Theoretical Physics, Department of Quantum and Classical Dynamics, Johannes Kepler Universit\"at Linz, Austria
}%

\author{David Gelbwaser-Klimovsky}
\affiliation{%
Schulich Faculty of Chemistry and Helen Diller Quantum Center, Technion-Israel Institute of Technology, Haifa 3200003, Israel}%

\date{\today}

\begin{abstract}
    Quantum heat transport for non-equilibrium steady state (NESS) exhibits
    a characteristic turnover effect, where the heat current reaches a maximum
    and subsequently declines as system-bath coupling increases. Although
    numerically exact methods can simulate this non-monotonic behavior, they
    offer limited information on the thermal state of the heat baths. Here, we introduce a
    frequency-selective thermometric protocol to probe the baths sustaining an
    NESS. By extracting a frequency-resolved effective temperature spectrum
    using a tunable two-level probe, we demonstrate that spectral dispersion
    serves as a direct witness for the non-equilibrium state of the heat baths.
    To demonstrate the protocol, we applied the hierarchical equations of motion
    to spin-boson and two-qubit models, though any exact method can be used. For
    both models, the turnover effect can be explained by how the thermal state of
    the heat baths evolves as the system-bath coupling strength increases.
\end{abstract}

\maketitle

\section{Introduction}
Understanding and controlling heat flow at the nanoscale is a fundamental
objective of quantum thermodynamics and device
engineering.\cite{cahillNanoscaleThermalTransport2003, Segal2003,
dubiColloquiumHeatFlow2011} As technology moves towards the realization of
efficient quantum thermal machines, mapping how heat transport scales with
system and bath parameters becomes important. A particularly intriguing
steady-state phenomenon in this domain is the turnover effect:\cite{
    Nicolin2011a, segalHeatTransferSpinboson2014,
    velizhaninMeirWingreenFormula2010, Yang2014, Gelbwaser-klimovsky2015a, 
wangnonequilibriumEnergyTransfer2015, wangUnifyingQuantumHeat2017,
Anto-sztrikacs2022, Anto-sztrikacs2023, Velizhanin2008, Boudjada2014,
saitoKondoSignatureHeat2013, Kato2015, Song2017, Pleasance2024} as the coupling
strength between a quantum system and thermal reservoirs increases, the
steady-state heat current initially rises, reaches a maximum at a critical
coupling strength, and subsequently declines. This non-monotonic behavior
reflects a fundamental transition in transport regimes, yet a clear physical
picture of the environment’s internal state during this process has remained
elusive.

Historically, the theoretical study of these systems has been split between
analytical tractability and numerical exactness.\cite{Anto-sztrikacs2023}
Traditional perturbative master equations, such as the standard Redfield
equation,\cite{Breuer2002, Segal2005} are widely used to study open quantum
systems due to their analytical simplicity. However, these methods fail to
capture the turnover effect entirely.\cite{Velizhanin2008, Kato2015} While more
sophisticated approximations--such as the non-interacting blip approximation
(NIBA),\cite{Nicolin2011a, segalHeatTransferSpinboson2014} Green’s function
techniques,\cite{velizhaninMeirWingreenFormula2010, Yang2014}
polaron-transformed approaches,\cite{Gelbwaser-klimovsky2015a, wangnonequilibriumEnergyTransfer2015,
wangUnifyingQuantumHeat2017} or the reaction coordinate mapping based
approaches\cite{Anto-sztrikacs2022} provide qualitative insight, they still
neglect the complex internal changes of the environment in a non-equilibrium
steady state (NESS).

Conversely, numerically exact methods like the multi-configurational
time-dependent Hartree (MCTDH) approach,\cite{Velizhanin2008} quantum Monte
Carlo simulations,\cite{saitoKondoSignatureHeat2013} path-integral
techniques,\cite{Boudjada2014} or the hierarchical equations of motion
(HEOM)\cite{Kato2015, Song2017, Pleasance2024} can simulate the turnover with
high fidelity, but they often function as ``black boxes'' regarding the bath's
response, offering little transparency into the local thermal conditions that
sustain and--in the strong coupling limit--suppress the heat current.

Here, we introduce a non-invasive thermometric protocol to quantify the thermometric
state of baths sustaining a NESS. Our approach utilizes a frequency-selective
probe\cite{Alicki2015, Pawutinan2025} to extract a ``local'' effective temperature
of the baths. By scanning the probe's transition frequency, we extract a
frequency-resolved ``effective temperature spectrum'' of the bath. Unlike a
macroscopic bath in equilibrium, which possesses a singular global temperature,
a non-equilibrium environment reveals its nature through spectral dispersion:
while in equilibrium the spectrum collapses to a constant, any variation across
frequencies serves as a signature for the non-equilibrium state. 
In particular, we find that in the strong coupling limit, the effective temperature approaches the initial temperature of the baths. This indicates an effective decoupling from the system, thereby explaining the turnover effect.

We demonstrate this protocol using the HEOM approach, applied to two canonical
models of quantum transport: the non-equilibrium spin-boson (NESB) model and a
two-qubit system where each spin is coupled to an independent reservoir.
Although showcased via HEOM, the protocol is platform-agnostic and can be
integrated into any exact numerical method.

The remainder of this paper is organized as follows. We discuss the theory in
Sec.~\ref{sec: theory}. We start by recapitulating the HEOM
approach in Sec.~\ref{sec: HEOM} and introduce the two models in Sec.~\ref{sec:
models}. The thermometric protocol is presented in Sec.~\ref{sec: effective
temperature} and a thermodynamically consistent definition of the heat current
is discussed in Sec.~\ref{sec: heat current}. The resulting effective
temperature spectra and heat current curves are are presented in Sec.~\ref{sec:
results}. Finally, Sec.~\ref{sec: conclusion} concludes the results.

\section{Model and Methodology}
\label{sec: theory}
The total Hamiltonian for a general model of an open quantum system with $K$
baths is given by
\begin{equation}\label{eqn: total hamiltonian}
	\hat{H}_\mathrm{tot} = \hat{H}_\mathrm{S} + \sum_{k=1}^K \left(
    \hat{H}_\mathrm{SB}^{(k)}+\hat{H}_\mathrm{B}^{(k)}
        \right)
\end{equation}
where $\hat H_\mathrm{S}$ is the system Hamiltonian, and $\hat
H_\mathrm{B}^{(k)}$ and $\hat H_\mathrm{SB}^{(k)}$ denote, respectively, the
Hamiltonian of the bath $k$ and its interaction with the system.

Each reservoir is modeled, like in the standard Caldeira-Leggett framework, as
an infinite set of non-interacting harmonic oscillators. The Hamiltonian of
bath $k$ is
\begin{equation*}
	\hat{H}_\mathrm{B}^{(k)} = \sum_{j} \hbar \omega_{k_j}
	\hat{b}_{k_j}^{\dagger} \hat{b}_{k_j}
\end{equation*}
where $\hat b_{k_j}$ ($\hat b_{k_j}^\dagger$) annihilates (creates) a quantum
in the mode $j$ of the bath $k$ with frequency $\omega_{k_j}$. The system-bath
interaction is taken to be bilinear in system and bath coordinates,
\begin{equation}\label{eqn: system-bath interaction}
    \hat{H}_\mathrm{SB}^{(k)} = \sum_{j} g_{k_j} \hat{V}_k \left(
	\hat{b}_{k_j}^{\dagger}+\hat{b}_{k_j}
        \right)
\end{equation}
where the Hermitian system operator $\hat V_k$ selects the degree(s) of freedom
of the system that is (are) coupled to the bath $k$, and $g_{k_j}$ is the
coupling constant for the mode $j$ of the heat bath. To account for the shift in the minimum of each oscillator
potential caused by the system-reservoir coupling, we include the standard
counter-term~\cite{Breuer2002, weissQuantumDissipativeSystems2001}
so that the total Hamiltonian
$
\hat H_\mathrm{tot} \to
    \hat H_\mathrm{S}(t) +
    \sum_k\big(
        \hat H_\mathrm{B}^{(k)} +
        \hat H_\mathrm{SB}^{(k)} +
        \hat H_\mathrm{ct}^{(k)}
    \big)
$ with
\begin{equation}\label{eqn: counter term}
    \hat H_\mathrm{ct}^{(k)}=\left(
    \hat V_\mathrm{S}^{(k)}\right
    )^{2} \frac{1}{\pi}\int_{0}^{\infty}d\omega \frac{\mathcal J_k(\omega)}{\omega}
\end{equation}
where spectral density
$
    \mathcal J_k(\omega)=\pi\sum_{j} |g_{k_j}|^2 \delta\big(
    \omega-\omega_{k_j}
    \big)
$
specifies both the mode distribution and the coupling strengths. In practical
applications, we take a continuous spectral
density\cite{weissQuantumDissipativeSystems2001, Breuer2002,
mayChargeEnergyTransfer2011} instead of a discrete sum of modes and a single
parameter--say $\lambda_k$ for $k^{th}$ bath--is typically factored out to
represent the ``overall'' coupling strength. Note that some
authors\cite{Pleasance2024, Anto-sztrikacs2021} adopt a convention in which the
spectral density is proportional to $\lambda_k^2$.

Assuming that the system and the $K$ independent heat baths are initially decoupled,
the total initial state is given by the product state $\hat\rho(0) =
\hat\rho_\mathrm{S}(0) \otimes \prod_{k} \hat\rho_\mathrm{B}^{(k)}$. Here, each
bath is prepared in a stationary thermal state:
\begin{equation}
    \hat\rho_\mathrm{B}^{(k)} = \frac{e^{-\beta_k \hat H_\mathrm{B}^{(k)}}}{\operatorname{Tr}\bigl\{e^{-\beta_k \hat H_\mathrm{B}^{(k)}}\bigr\}}
\end{equation}
where $\beta_k = (k_\mathrm{B}T^{(k)})^{-1}$ is the inverse temperature of the
heat bath $k$. 

Under these conditions, the collective bath coordinate $\hat X_k = \sum_{j}
g_{k_j}\bigl( \hat{b}_{k_j}^{\dagger}+ \hat{b}_{k_j} \bigr)$ exhibits Gaussian
statistics. Consequently, environmental influence is entirely determined by
the two-time bath correlation function $\mathcal C_k(t) =
\mathcal{C}_k^\mathrm{Re}(t) + i\mathcal{C}_k^\mathrm{Im}(t) = \langle \hat
X_k(t) \hat X_k(0)\rangle_\mathrm{B}$, where the expectation value is taken
with respect to the canonical density operator of the baths. This correlation
function can be explicitly evaluated as~\cite{Breuer2002,
Tanimura1989, Jin2008, Tanimura2020, Lambert2023}
\begin{equation}\label{eqn: bath correlation function}
\begin{aligned}
    \mathcal{C}_k(t) = \int_0^{\infty} &d \omega \frac{\mathcal{J}_k(\omega)}{\pi} \\
    &\times \left[ \coth\left(\frac{\beta_k \hbar \omega}{2}\right) \cos(\omega t) - i\sin(\omega t) \right]
\end{aligned}
\end{equation}

This correlation function enters the time-ordered influence functional, which
yields the exact time evolution of the reduced system density
matrix~\cite{Breuer2002,Tanimura1989, Ishizaki2005}: 
\begin{equation}\label{eqn: influence functional}
\begin{aligned}
    \tilde{\rho}_\mathrm{S}(t) = \mathcal{T}_+ \prod_{k=1}^K \exp \Biggl\{ 
    &-\frac{1}{\hbar^2}\int_0^t d t_2 \int_0^{t_2} d t_1\,\tilde{V}_k(t_2)^{\times} \\
    &\quad\Big[ \mathcal{C}_k^\mathrm{Re} (t_2-t_1) \tilde{V}_k(t_1)^{\times} \\
    &\quad + i \mathcal{C}_k^\mathrm{Im}(t_2-t_1) \tilde{V}_k(t_1)^{\circ} \Big] \Biggr\}
    \tilde{\rho}_\mathrm{S}(0)
\end{aligned}
\end{equation}
where $\mathcal{T}_+$ denotes the chronological time-ordering operator. Here,
the tilde $\tilde{O}(t)$ indicates an operator in the interaction picture with
respect to $\hat H_\mathrm{S}(t) + \hat H_\mathrm{B}$, while the superoperators
$\hat O^{\times}$ and $\hat O^{\circ}$ represent the commutator $[\hat O,
\cdot]$ and anti-commutator $\{\hat O, \cdot\}$, respectively.

Although the time-ordered influence functional [Eq.~\eqref{eqn: influence
functional}] provides a formally exact description of open system dynamics, its
non-local time integrals render analytical evaluation intractable. To enable
efficient numerical simulation, the Hierarchical Equations of Motion (HEOM)
formalism transforms the path integral into an infinite set of time-local
coupled differential equations.

\subsection{Hierarchical Equations of Motion}
\label{sec: HEOM}
The Hierarchical Equations of Motion (HEOM) formalism is a non-perturbative
approach originally introduced by \textcite{Tanimura1989} to simulate open
quantum systems coupled to non-Markovian environments at finite temperatures.
This method unravels the exact influence functional [Eq.~\eqref{eqn: influence
functional}] into a hierarchy of coupled differential
equations.\cite{Tanimura2020, Jin2008}

To implement the HEOM, the real and imaginary components of the bath
correlation functions are assumed to be expandable as a linear combination of exponential
functions. For $t \geq 0$, we define:
\begin{equation}\label{eqn: exp decomposition of BCF}
    \mathcal C_k^\mathrm{Re}(t) =
    \sum_{j=0}^{J_k^\mathrm{Re}} c_{k_j}^\mathrm{Re} e^{-\nu_{k_j}^\mathrm{Re} t}
    \quad \text{and} \quad
    \mathcal C_k^\mathrm{Im}(t) =
    \sum_{j=0}^{J_k^\mathrm{Im}} c_{k_j}^\mathrm{Im} e^{-\nu_{k_j}^\mathrm{Im} t}
\end{equation}
where the amplitudes $c_{k_j}^\alpha$ and frequencies $\nu_{k_j}^\alpha$
($\alpha \in \{\mathrm{Re}, \mathrm{Im}\}$) are generally complex-valued, and the number of exponential terms included in the decomposition is given by $J_k^\mathrm{Re}+1$ and $J_k^\mathrm{Im}+1$ for the real and imaginary parts, respectively. Although alternative
functional expansions have been proposed,\cite{Ikeda2020} this work adheres to
the standard exponential decomposition scheme. For
negative times ($t < 0$), the correlation functions follow the time-reversal
symmetry $\mathcal{C}_k(-t) = \mathcal{C}_k^*(t)$.

Substituting Eq.~\eqref{eqn: exp decomposition of BCF} into the influence
functional enables the mapping of the path integral to a set of time-local
differential equations. This is achieved by introducing a set of auxiliary
density operators (ADOs), denoted as $\hat\rho_{\mathbf{n}}(t)$. Here, the
global multi-index $\mathbf{n} = (\vec{n}_1, \ldots, \vec{n}_K)$ acts as a
matrix-label where each column vector $\vec{n}_k = (n_{k_0}^\mathrm{Re},
\ldots, n_{k_{J_k^\mathrm{Re}}}^\mathrm{Re}, n_{k_0}^\mathrm{Im}, \ldots,
n_{k_{J_k^\mathrm{Im}}}^\mathrm{Im})^T$ consists of non-negative integers. In
the Schrödinger picture, the resulting coupled differential equations take the
form:
\begin{equation}\label{eqn: HEOM time evolution}
    \begin{aligned}
        \frac{\partial}{\partial t}\hat\rho_{\mathbf{n}}(t) = 
        & \left(\mathcal{L}_\mathrm{S} - \sum_{k=1}^K \sum_{\alpha} \sum_{j=0}^{J_k^\alpha} n_{k_j}^\alpha \nu_{k_j}^\alpha \right) \hat\rho_{\mathbf{n}}(t) \\
        & -\frac{i}{\hbar} \sum_{k=1}^K \hat{V}_k^{\times} \sum_{j=0}^{J_k^\mathrm{Re}} n_{k_j}^\mathrm{Re} c_{k_j}^\mathrm{Re} \hat\rho_{\mathbf{n} - \vec{e}_{k,\mathrm{Re},j}}(t) \\
        & +\frac{1}{\hbar} \sum_{k=1}^K \hat{V}_k^{\circ} \sum_{j=0}^{J_k^\mathrm{Im}} n_{k_j}^\mathrm{Im} c_{k_j}^\mathrm{Im} \hat\rho_{\mathbf{n} - \vec{e}_{k,\mathrm{Im},j}}(t) \\
        & -\frac{i}{\hbar} \sum_{k=1}^K \hat{V}_k^{\times} \sum_{\alpha} \sum_{j=0}^{J_k^\alpha} \hat\rho_{\mathbf{n} + \vec{e}_{k,\alpha,j}}(t)
    \end{aligned}
\end{equation}
where $\mathcal{L}_\mathrm{S} = -\frac{i}{\hbar}\hat{H}_\mathrm{S}^\times$ is
the Liouvillian of the system, and $\vec{e}_{k,\alpha,j}$ represents the unit vector
that shifts (increments or decrements) the specific index component
$n^\alpha_{k_j}$ within the global multi-index $\mathbf{n}$.

Each ADO shares the same dimensionality as the reduced system density matrix.
The root of the hierarchy, where all indices vanish ($\mathbf{n}=\mathbf{0}$),
corresponds to the physical reduced density operator of the system, i.e.,
$\hat\rho_\mathrm{S}(t) = \hat\rho_{\mathbf{0}}(t)$. Higher-order ADOs
($\mathbf{n} \neq \mathbf{0}$) capture the non-Markovian memory effects and
system-bath correlations. As an uncorrelated initial state is assumed
[Eq.~\eqref{eqn: influence functional}], it corresponds to the initial state of
the hierarchy with the root ADO set to the initial density matrix of the
system, $\hat\rho_{\mathbf{0}}(0) = \hat\rho_\mathrm{S}(0)$, while all
remaining ADOs are set to zero.

Although Eq.~\eqref{eqn: HEOM time
evolution} is formally exact for an infinite hierarchy, numerical
implementation requires a truncation scheme. A standard approach restricts
the hierarchy to a maximum tier $N$, such that:
\begin{equation}
    N_\mathrm{total} = \sum_{k=1}^K
    \sum_{\alpha}
    \sum_{j=0}^{J_k^\alpha}
    n_{k_j}^\alpha \leq N
\end{equation}
The convergence of this truncation is verified by systematically increasing $N$
until the dynamics of the root density matrix $\hat\rho_\mathrm{S}(t)$
stabilizes within a specified numerical tolerance.

Several highly optimized software packages are available to solve these
equations. Notable examples include frameworks optimized for parallel
acceleration across multi-core CPU and GPU architectures,\cite{Kreisbeck2011,
Strumpfer2012} as well as DM-HEOM,\cite{Noack2018, Kramer2018e} which leverages
distributed-memory structures across high-performance compute nodes to handle
exceptional hierarchy depths. In this work, we utilize the implementation of
HEOM within the QuTiP framework.\cite{Johansson2012, Johansson2013,
Lambert2023, Lambert2024} This choice provides a versatile, general-purpose
solver well-suited for our diverse parameter regime, offering a flexible
alternative to highly specialized tools designed strictly for specific system
topologies.

In this work, we focus exclusively on the asymptotic long-time limit where the
composite system with baths at different temperatures has relaxed into an NESS,
defined by the condition
\begin{equation}\label{eqn: steady state condition}
\partial\hat\rho_{\mathbf{n}}^\mathrm{ss} / \partial t = 0
\end{equation}
for all multi-indices $\mathbf{n}$. We chose the following models for their computational simplicity. Both of them show the turnover effect.

\subsection{Demonstrative Models}\label{sec: models}
To demonstrate the thermometry protocol, we analyzed the following two models

\begin{figure}
    \centering
    \includegraphics[]{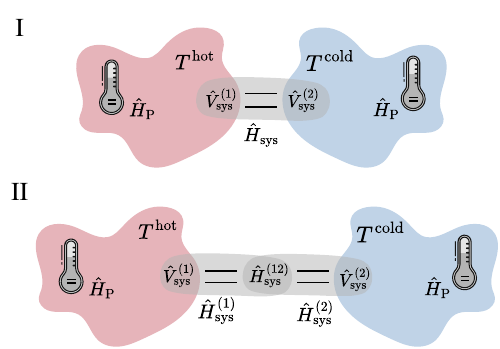}
    \caption{
        Schematic of (I) the non-equilibrium spin-boson model and 
        (II) model of two interacting qubits coupled to separate heat baths.
        Baths in both models are probed with two level thermometers.
    }
    \label{fig: schematic with thermometers}
\end{figure}

\textbf{Model I:} A single two-level system (spin) simultaneously coupled to
two independent harmonic (bosonic) reservoirs at different
temperatures~[Fig.~\ref{fig: schematic with thermometers}-(I)]. The system
Hamiltonian is given by
$$\hat{H}_\mathrm{sys} = \frac{\hbar\omega_0}{2}\,\hat{\sigma}_z$$
where $\hbar\omega_0$ is the energy difference between the two
levels. Spin is coupled via the interaction operator $\sigma_x$ for both baths.

\textbf{Model II:} Two interacting spins, each coupled to an independent heat
bath, with the two baths at different initial temperatures~[Fig.~\ref{fig:
schematic with thermometers}-(II)]. The system Hamiltonian
$\hat{H}_\mathrm{sys} = \hat H_\mathrm{sys}^{(1)} +\hat H_\mathrm{sys}^{(2)} +
\hat H_\mathrm{sys}^{(12)}$ has the individual qubit Hamiltonian, and
qubit-qubit interaction Hamiltonian given by
\begin{equation*}
\begin{aligned}
    \hat H_\mathrm{sys}^{(m)} &= \frac{\hbar\omega^{(m)}_0}{2} (\hat{\sigma}_z^{(m)}
    + \mathbb{I}_2^{(m)})\\
    \hat H_\mathrm{sys}^{(12)} &= \hbar J_{12} \big( \hat{\sigma}_+^{(1)}\hat{\sigma}_-^{(2)}
              + \hat{\sigma}_-^{(1)}\hat{\sigma}_+^{(2)} \big)
\end{aligned}
\end{equation*}
where $\omega_0^{(m)}$ is the bare frequency of the $m$-th spin and $J_{12}$
represents the coherent inter-spin coupling strength. Each spin is coupled with its bath with the interaction operator $\sigma_x$.

For both models, we chose the heat baths with the Drude-Lorentz spectral density~\cite{Ritschel2014, Lambert2023}
\begin{equation}\label{eqn: Drude-Lorentz spectral density}
    \mathcal{J}_k(\omega) = \frac{2 \lambda_k \gamma_k \omega}
                       {\left(\gamma_k^2+\omega^2\right)}
\end{equation}
where $\lambda_k$ represents the coupling strength and $\gamma_k$ is the
Lorentzian cut-off frequency for the heat bath $k$. 

A fundamental assumption in the derivation of the hierarchical equations of
motion is the exponential decomposition of the bath correlation functions
[Eq.~\eqref{eqn: exp decomposition of BCF}]. For the Drude-Lorentz spectral
density, this decomposition can be calculated analytically. See
appendix~\ref{app: pade decomposition} for an efficient decomposition and a
terminator correction in the system Liouvillian due to truncation.

In both models, each heat bath is coupled to a two-level thermometer, referred to as the probe. The probe's degrees of freedom are integrated into the total Hilbert space through:
\begin{equation} 
    \hat{H}_\mathrm{tot} = \hat{H}_\mathrm{sys} 
                             + \sum_{k\in\{\mathrm{hot, cold}\}} \left(
                                 \hat{H}_\mathrm{P}^{(k)} +
                                 \hat{H}_\mathrm{B}^{(k)} +
                                 \hat{H}_\mathrm{sysB}^{(k)} +
                                 \hat{H}_\mathrm{PB}^{(k)} +
                                 \hat{H}_\mathrm{ct}^{(k)}
                         \right)
\end{equation} 
where the probe with Hamiltonian $H_\mathrm{P}^{(k)}$ is coupled to the bath,
with the coupling Hamiltonian $\hat{H}_\mathrm{PB}^{(k)}$. 

For simplicity, we consider that the probe-bath interaction is bilinear in the
probe and bath coordinates with the same spectral density as the system-bath
interaction but with a coupling strength $\eta\ll\lambda_k$. It can be achieved
by 
\begin{equation}\label{eqn: probe-bath interaction}
    \hat{H}_\mathrm{PB}^{(k)} = \sum_{j} g_{k_j}\sqrt{\frac{\eta}{\lambda_k}}\,\hat{V}_P^{(k)} \left(
	\hat{b}_{k_j}^{\dagger}+\hat{b}_{k_j}
        \right)
\end{equation}
Hence, the spectral density for the probe interaction with the heat bath $k$ is
given by
$
    \mathcal J_k^{(P)}(\omega)=\pi\sum_{j} \frac{\eta}{\lambda_k}|g_{k_j}|^2 \delta\big(
    \omega-\omega_{k_j}
    \big)
$.
This can be incorporated into the interaction operator of the composite system
corresponding to the bath $k$
\begin{equation}
    \label{eqn: interaction operator with probe}
    \hat{V}_{k} = \hat{V}_\mathrm{S}^{(k)} \otimes
                         \mathbb{I}_\mathrm{P} +
                         \sqrt{\frac{\eta}{\lambda_{k}}}\,\mathbb{I}_\mathrm{S} \otimes
                         \hat{V}_\mathrm{P}
\end{equation}
The operator $\hat{V}_\mathrm{S}^{(k)}$ and $\hat{V}_\mathrm{P}^{(k)}$ act
exclusively on the system and the probe respectively. The spectral scaling
$\eta$ is kept perturbative relative to the intrinsic system and bath energy
scales to ensure minimal back action. The corresponding counter-term
$\hat{H}_\mathrm{ct}^{(k)}$ takes the standard quadratic form in
$\hat{V}_k$~[Eq.~\eqref{eqn: counter term}].

We used two level probes with Hamiltonian $\hat{H}_\mathrm{P} =
{\hbar\omega}/{2}\,\hat\sigma_z$ and exchange energy with their
respective heat baths via the coupling operator $\hat V_\mathrm{P} =
\hat\sigma_x$.

For numerical efficiency, we simulate the system with a single probe attached at a time; this approach yields identical results to a full two-probe simulation.

\subsection{Effective Temperature via Bath-Coupled Probe}
\label{sec: effective temperature}
Temperature is fundamentally an equilibrium state variable. For a bath away
from equilibrium, a global temperature is generally undefined; nevertheless,
one can assign a frequency-dependent local temperature using a minimally
invasive thermometric probe.\cite{Alicki2015, Pawutinan2025} In the steady
state limit of the bath, the energy eigenstate population of a weakly coupled
two-level probe of transition frequency $\omega$ would satisfy the following
detailed balance relation
\begin{equation}\label{eqn: effective temperature definition}
    \frac{P_\mathrm{e}^\mathrm{ss}}{P_\mathrm{g}^\mathrm{ss}} =
    \exp\left[-\frac{\hbar\omega}{k_\mathrm{B} T_\mathrm{eff}(\omega)}\right],
\end{equation}
where $P_\mathrm{e}^\mathrm{ss}$ and $P_\mathrm{g}^\mathrm{ss}$ denote
the excited- and ground-state populations of the probe respectively.
Therefore, we define the frequency-dependent effective temperature as
\begin{equation}
    T_\mathrm{eff}(\omega) = \frac{\hbar\omega}{k_\mathrm{B}\ln\left(
    P_\mathrm{g}^\mathrm{ss}/P_\mathrm{e}^\mathrm{ss}
    \right)}
\end{equation} 
In global equilibrium, $T_\mathrm{eff}(\omega)$ becomes independent of $\omega$
and is the physical bath temperature; departures from flatness across the
frequency spectrum is a signature of non-equilibrium bath state.

For a probe of transition frequency $\omega$, which is coupled to the heat bath
$k$, the operational workflow is structured as follows:
\begin{itemize}
    \item \textbf{Steady-State Extraction:} We find the steady-state of HEOM by
        solving for the null-space of HEOM~[Eq.~\eqref{eqn: steady state
        condition}] using QuTiP. The steady-state populations
        $P_\mathrm{e,g}^\mathrm{ss}$ are extracted from the reduced probe
        density matrix to compute $T_\mathrm{eff}^{(k)}(\omega)$ by
        Eq.~\eqref{eqn: effective temperature definition}.
    \item \textbf{Minimal Invasiveness:} Because the probe is weakly
        coupled to the infinite bath and is small compared to it, we would
        expect minimal invasiveness. Which is verified by confirming that bare
        system observables and inter-reservoir heat currents remain invariant
        within numerical tolerance as probe-bath coupling ($\eta$) is varied.
    \item \textbf{Spectrum Generation:} The protocol is swept systematically
        over a range of transition frequencies $\omega$ to construct the
        reservoir’s temperature spectrum $T_\mathrm{eff}^{(k)}(\omega)$. A flat
        spectrum is a signature of  near-equilibrium conditions, whereas strong
        frequency dependence signals local non-equilibrium structures.
\end{itemize}
Hence this purely bath-coupled thermometry protocol provides a minimally
invasive diagnostic of individual reservoirs.

\subsection{Heat Current Evaluation}\label{sec: heat current}
To provide a thermodynamic interpretation and verify the consistency of our
effective temperature results, we monitor the energy exchange between the
system and the environments. Following Kato and
Tanimura,\cite{katoHierarchicalEquationsMotion2018, Kato2016} we adopt a
thermodynamically consistent definition for the heat current entering from the
$k$-th bath:
\begin{equation}\label{eqn: heat current definition}
    \dot{Q}_\mathrm{B}^{(k)}(t) \equiv - \frac{d}{dt} \bigl\langle
                                    \hat{H}_\mathrm{B}^{(k)}(t)
                                    \bigr\rangle
\end{equation}
Assuming that both the isolated bath Hamiltonian $\hat{H}_\mathrm{B}^{(k)}$
and the system-bath interaction terms are time-independent in the Schrödinger
picture, this energy flow can be evaluated directly within the HEOM framework
using the first-tier auxiliary density operators. The explicit evaluation
yields
\begin{equation}\label{eqn: heat current HEOM}
    \begin{aligned}
        \dot{Q}_\mathrm{B}^{(k)}(t) = 
        & -\sum_{\alpha \in \{\mathrm{Re,Im}\}} \sum_{j=0}^{J_k^\alpha} \nu_{k_j}^\alpha \operatorname{Tr}\bigl\{ \hat{V}_k \hat{\rho}_{\vec{e}_{k,\alpha,j}}(t) \bigr\} \\
        & + \frac{2}{\hbar} \mathcal{C}_k^\mathrm{Im}(0) \operatorname{Tr}\bigl\{ \hat{V}_k^2 \hat{\rho}_\mathrm{S}(t) \bigr\} \\
        & + \frac{i}{\hbar} \Delta_k \operatorname{Tr}\bigl\{ \bigl[ \hat{A}_k(t), \hat{V}_k \bigr] \hat{\rho}_\mathrm{S}(t) \bigr\} \\
        & + \Delta_k \sum_{k^{\prime} \neq k} \sum_{\alpha \in \{\mathrm{Re,Im}\}} \sum_{j=0}^{J_{k^{\prime}}^\alpha} \operatorname{Tr}\bigl\{ \hat{B}_{k, k^{\prime}} \hat{\rho}_{\vec{e}_{k^{\prime},\alpha,j}}(t) \bigr\} \\
        & + \frac{i}{\hbar} \Delta_k \sum_{k^{\prime} \neq k} \Delta_{k^{\prime}} \operatorname{Tr}\bigl\{ \bigl[ \hat{B}_{k, k^{\prime}}, \hat{V}_{k^{\prime}} \bigr] \hat{\rho}_\mathrm{S}(t) \bigr\}
    \end{aligned}
\end{equation}
where, $\hat A_k = {i}/{\hbar}[\hat H_\mathrm{S}, \hat V_k]$ and $B_{k,
k^\prime} = (i/\hbar)^2[[\hat V_k, \hat V_{k^\prime}], \hat V_k]$. Here,
$\hat{\rho}_{\vec{e}_{k,\alpha,j}}(t)$ denotes the first-tier ADO where only
the single index corresponding to the $j$-th mode of the $\alpha$-component of
the $k$-th bath is set to one, with all other indices zero. The terminator, $\Delta_k$, (see Appendix~\ref{app: pade decomposition}) correction for the
heat current is given by the last three terms.

\section{Results and Discussion}
\label{sec: results}

\begin{figure}
    \centering
    \includegraphics[]{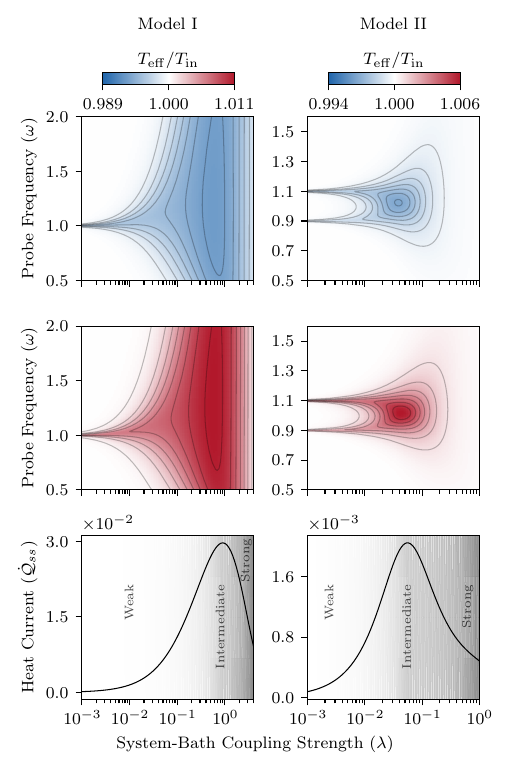}
    \caption{
        Units: $\hbar = k_\mathrm{B} = 1$. Left column: Model I
        (non-equilibrium spin-boson) with system transition frequency $\omega_0
        = 1$. Right column: Model II (two interacting qubits) with individual
        frequencies $\omega_0^{(1)} = \omega_0^{(2)} = 1$ and inter-qubit
        coupling $J_{12} = 0.1$. In both models, the two independent reservoirs
        have Drude–Lorentz spectral density ($\gamma_\mathrm{hot} = \gamma_\mathrm{cold} = 1$ for Model I; $\gamma_\mathrm{hot} = \gamma_\mathrm{cold} = 5$ for Model II) at initial temperatures
        $T_\mathrm{in}^\mathrm{hot} = 3$ and $T_\mathrm{in}^\mathrm{cold} = 2$ and $\lambda_\mathrm{hot}=\lambda_\mathrm{cold}=\lambda$.
        The embedded two-level thermometers are coupled weakly with coupling
        strength $\eta = 10^{-8}$. The first two rows show
        $T_\mathrm{eff}(\omega,\lambda)$ normalized by the respective initial
        bath temperature $T_\mathrm{in}$ (top: hot bath; second: cold bath).
        Probes read below $T_\mathrm{in}$ on the hot side and above
        $T_\mathrm{in}$ on the cold side at weak to intermediate coupling. In
        the strong-coupling regime, $T_\mathrm{eff}(\omega,\lambda)$ across
        frequencies converges toward a common asymptotic value. Bottom row: Steady-state heat current, where $\dot{Q}_\mathrm{ss} \equiv \dot{Q}_\mathrm{B}^\mathrm{hot}(t\rightarrow \infty) = -\dot{Q}_\mathrm{B}^\mathrm{cold}(t\rightarrow \infty)$.
For the hierarchical equations of motion calculations, the number of
        Pad\'e frequencies were 2 ($J_k=1$) for Model I and 3 ($J_k=2$) for Model II.
        }
    \label{fig: results}
\end{figure}
\begin{figure}[]
    \centering
    \includegraphics[]{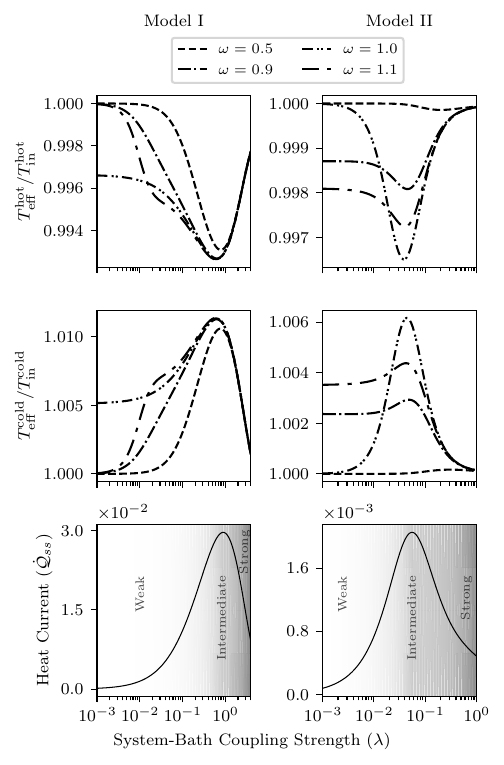}
    \caption{Effective temperature as a function of coupling strength ($\lambda$) for representative probe frequencies $\omega$ (sliced from data in Fig.~\ref{fig: results}).}
    \label{fig: result_slice}
\end{figure}

Applying the frequency-resolved thermometry protocol (Sec.~\ref{sec: effective temperature}) to the non-equilibrium steady states of both models revealed three distinct transport regimes as the system-bath coupling was increased [Fig.~\ref{fig: results}]. Notably, the regime boundaries differ between the two models, demonstrating that these transitions are not governed solely by the spectral density coupling strength ($\lambda$), but also depend on other system and bath parameters.

\paragraph{Weak-coupling regime.} In the weak-coupling regime, probes with frequencies near the allowed eigenenergy transitions of the bare system exhibited peak temperature deviations, whereas off-resonant probes reported values closer to the initial bath temperatures. This behavior is consistent with resonant energy transport in the weak-coupling limit.\cite{Segal2005, Segal2006}

For Model~I, the characteristic energy gap was set to $\hbar\omega_0=1$ [Fig.~\ref{fig: results} - Model I]; varying $\omega_0$ accordingly shifted the peak. 

For Model~II, the energy eigen-system is given by:
\begin{equation*}
\begin{aligned}
E_{\lvert\downarrow\downarrow\rangle} &= 0, & 
E_{\frac{1}{\sqrt{2}}(\lvert\uparrow\downarrow\rangle - \lvert\downarrow\uparrow\rangle)} &= 0.9, \\
E_{\frac{1}{\sqrt{2}}(\lvert\uparrow\downarrow\rangle + \lvert\downarrow\uparrow\rangle)} &= 1.1, & 
E_{\lvert\uparrow\uparrow\rangle} &= 2.0.
\end{aligned}
\end{equation*}
Transforming the inter-system interaction operator, $\hat{H}_\mathrm{sys}^{(12)} = J_{12} \big( \hat{\sigma}_+^{(1)}\hat{\sigma}_-^{(2)} + \hat{\sigma}_-^{(1)}\hat{\sigma}_+^{(2)} \big)$, into the energy eigenbasis reveals that the allowed transitions are restricted to
\begin{equation*}
\lvert\downarrow\downarrow\rangle \rightleftharpoons \frac{1}{\sqrt{2}}\big(\lvert\uparrow\downarrow\rangle \mp \lvert\downarrow\uparrow\rangle\big) \quad \text{and} \quad \frac{1}{\sqrt{2}}\big(\lvert\uparrow\downarrow\rangle \mp \lvert\downarrow\uparrow\rangle\big) \rightleftharpoons \lvert\uparrow\uparrow\rangle.
\end{equation*}
Thus, the energy gaps for the allowed transitions were $0.9$ and $1.1$, which correspond to the observed peaks~[Fig.~\ref{fig: results} - Model II].

\paragraph{Intermediate-coupling regime.} As the coupling was increased, the peaks broadened and shifted away from the bare transition frequencies. The spectral dispersion of the effective temperature reflects how far the bath state departs from thermal equilibrium. Moreover, the increased spectral dispersion coincided with enhanced energy transport, which our protocol showed occurs via non-resonant channels.

\paragraph{Strong-coupling regime.} In the strong-coupling regime, effective temperatures measured across different probe frequencies approached a single value of initial bath temperature. This frequency-by-frequency agreement together with the reduction of temperature deviations suggests that in the strong coupling limit, the relevant system transitions are effectively decoupled from the baths, leading to the suppression of steady-state energy exchange.

For  clarity, $T_\mathrm{eff}/T_\mathrm{in}$ with respect to the coupling strength is shown for a few representative probe frequencies in Fig.~\ref{fig: result_slice}.

\section{Conclusion}\label{sec: conclusion}
By providing an operational window into the bath's internal state beyond the weak-coupling limit, frequency-resolved thermometry offers a new lens through which to view the turnover effect. Our findings demonstrate how the effective temperature for the different frequency modes of the bath evolves with coupling strength. Specifically, the frequency dispersion of the effective temperature serves as a quantitative measure of the bath's deviation from thermal equilibrium, directly correlating with the rise and fall of the heat current.

Our results established three distinct operational regimes. The weak-coupling regime was characterized by resonant energy transfer concentrated at the bare system transition frequencies. In the intermediate-coupling regime, the resonance peaks broadened and shifted, accompanied by a maximum in the frequency dispersion of the effective temperature. In contrast, in the strong-coupling limit, this frequency dispersion diminished and the effective temperatures converged toward the initial bath values. This behavior provides a clear signature of effective bath decoupling, underlying the physical origin of the turnover effect.

Although the quantum Zeno effect (QZE)\cite{Yang2014, Kato2015} or the reduced transfer rate due to system-bath hybridization\cite{oehrlMulticavityStrongCoupling2026} have been proposed as an explanation for the turnover effect, systematically testing and validating these hypotheses remains an open challenge. Our work introduces a distinct operational mechanism that may complement these explanations.

Moreover, our results also shed light on the deviation of the bath state from the initial equilibrium state.   The deviation from equilibrium arises from the interaction with the system. Although it can be neglected in the weak-coupling limit, our results show that it plays a key role in the non-equilibrium dynamics at intermediate coupling. In this regime, the frequency dispersion of the effective temperature reaches its maximum, and therefore the bath state is at the "greatest distance'' from equilibrium. Neglecting changes in the bath state in this regime could yield incorrect results.  Surprisingly enough, this deviation decreases in the strong-coupling limit, where the effective temperature dispersion tends to disappear. This may suggest the possibility of neglecting changes in the bath state at this limit.

\section*{Acknowledgments}
We acknowledge funding from the Austrian Science Foundation through FWF
Project No. P35844, ``Open Quantum Dynamics Lab.''

\section*{Author Declarations}
\subsection*{Conflict of interest}
The authors have no conflicts of interest to disclose.
\subsection*{Author Contributions}
All authors contributed equally to this work.

\appendix
\section{Pad\'e Decomposition and Terminator for the Drude-Lorentz Spectral Density}
\label{app: pade decomposition}
This work uses the Padé decomposition,\cite{Hu2010, Hu2011} chosen for its
significantly faster convergence compared to the Matsubara
decomposition~\cite{Shi2009}.

In the Padé scheme presented below, the real and imaginary parts are not
treated with separate indices, unlike the general form in Eq.~\eqref{eqn: exp
decomposition of BCF}. This simplification is possible because the only term
with a non-zero imaginary component corresponds to $j=0$ for which
$\nu_{k_0}^\mathrm{Im}=\nu_{k_0}^\mathrm{Re}=\gamma_k$. Therefore, the
coefficients can be combined into a single index for a gain in numerical
efficiency~\cite{Fruchtman2016, Lambert2023}.

\begin{equation*}
    \mathcal{C}_k(t)=\sum_{j=0}^{\infty} c_{k_j} e^{-\nu_{k_j} t}
\end{equation*}
The expansion has infinite terms, but the numerical implementation of HEOM
(Eq.~\eqref{eqn: HEOM time evolution}) needs a finite decomposition, i.e. $j \leq
J_k$. Therefore, a truncation is done based on the approximation that if $1 /
\nu_{k_j}$ is much smaller than other important time-scales then
$e^{-\nu_{k_j} t} \approx \delta(t) / \nu_{k_j}$
\begin{equation}\label{eqn: finite pade BCF}
    \mathcal{C}_k(t) \approx \sum_{j=0}^{J_k} c_{k_j} e^{-\nu_{k_j} t} +
    \sum_{j=J_k+1}^{\infty} \frac{c_{k_j}}{\nu_{k_j}} \delta(t)
\end{equation}
thus, reducing the infinite sum of exponentials into a finite sum of
exponentials and infinite sum which captures the divergent real part at $t = 0$
which can be treated by terminator formalism provided in~\cite{Ishizaki2005}.
The Pad\'e decomposition parameters for the finite sum are given by
\begin{equation}
    \nu_{k_j}= \begin{cases}\gamma_k & j=0 \\
    \xi_{k_j}/(\beta_k\hbar) & j \geq 1\end{cases}
\end{equation}
\begin{equation}
    c_{k_j} = \begin{cases}
        \lambda_k \gamma_k \bigl[
            \cot (\beta_k \hbar \gamma_k / 2)-i
            \bigr] & j=0 \\
        4 \lambda_k \gamma_k {\eta_{k_j}\xi_{k_j}}\Big/{\left(
        \xi_{k_j}^2-\left(\gamma_k\beta_k\hbar\right)^2
        \right)} & j \geq 1
    \end{cases}
\end{equation}
where, $\xi_{k_j}$ and $\eta_{k_j}$ are obtained by diagonalizing two
specific matrices. The first matrix, $\Lambda$, is defined for
$m,n\in\{1,\ldots, 2 J_k\}$ with elements:
$$
    \Lambda_{m, n}=\frac{\delta_{m, n-1}}{\sqrt{(2 m+1)(2 n+1)}}+
    \frac{\delta_{m, n+1}}{\sqrt{(2 m+1)(2 n+1)}}
$$
Let its eigenvalues, arranged in ascending order, be denoted $\{\lambda_i\}$.
The second matrix, $\Lambda^\prime$, is defined for $m,n\in\{1,\ldots, 2
J_k-1\}$ as:
$$
\Lambda^\prime_{m, n} = \frac{\delta_{m, n-1}}{\sqrt{(2 m+3)(2 n+3)}} +
                        \frac{\delta_{m, n+1}}{\sqrt{(2 m+3)(2 n+3)}}
$$
Its ascending eigenvalues are denoted $\{\lambda^\prime_i\}$.
Then $\xi_{k_j}=-2/\lambda_j$ for $j\in\{1,\ldots,J_k\}$ and
$\zeta_{k_j}=-2/\lambda^\prime_j$ for $j\in\{1,\ldots,J_k-1\}$ and
$$
    \eta_{k_j} = \left(
                    J_k^2+\frac{3}{2} J_k
                    \right) \frac{\prod_{i=0}^{J_k-1}\left(
                    \zeta_{k_i}^2-\xi_{k_i}^2
                    \right)}{\prod_{i=1}^{J_k}\left(
                    \xi_{k_j}^2-\xi_{k_i}^2 + \delta_{i, j}
                    \right)}, \quad j\in\{1, \ldots, J_k\}
$$
The infinite sum in Eq.~\eqref{eqn: finite pade BCF} can be calculated by
taking the whole sum $\sum_{j=0}^{\infty}\frac{c_{k_j}}{\nu_{k_j}}$ and
subtracting the finite sum $\sum_{j=0}^{J_k}\frac{c_{k_j}}{\nu_{k_j}}$ from it.
The whole sum should be invariant for the type of decomposition used, and it
has been evaluated using the Matsubara decomposition~\cite{Lambert2023}
$$
    \sum_{j=0}^{\infty} \frac{c_{k_j}}{\nu_{k_j}} =
    \frac{2 \lambda_k}{\beta_k \hbar \gamma_k}-i \lambda_k
$$
It is then possible to show~\cite{Ishizaki2005} that the contribution of the
delta-function to the correlation functions can be described in the
HEOM~(Eq.~\eqref{eqn: HEOM time evolution}) by modifying
$\mathcal{L}_\mathrm{S}$ as
\begin{equation}
    \mathcal{L}_\mathrm{S} \rightarrow \mathcal{L}_\mathrm{S}+
    \frac{1}{\hbar^2}\sum_{k=1}^K \Delta_k \hat V_k^\times \hat V_k^\times
\end{equation}
where
\begin{equation}
    \Delta_k = \frac{2 \lambda_k}{\hbar \beta_k \gamma_k}-i \lambda_k -
    \sum_{j=0}^{J_k} \frac{c_{k_j}}{\nu_{k_j}}
\end{equation}

\end{document}